\documentstyle[trans]{rs}
\input epsf.tex
\begin{document}
\title[Strongly disordered magnetic systems]{Magnetic properties
of strongly disordered electronic systems}
\author[S. Sachdev]{Subir Sachdev}
\affiliation{Department of Physics, Yale University, P.O. Box 208120,\\
New Haven CT 06520-8120, USA.}
\maketitle
\label{firstpage}
\begin{abstract}
We present a unified, global perspective on the magnetic properties of strongly
disordered electronic systems, with special emphasis on the
case where the ground state is
metallic. We review the arguments for the instability of the disordered Fermi
liquid
state towards the formation of local magnetic moments, and argue that their
singular low temperature thermodynamics are
the ``quantum Griffiths'' precursors of the quantum phase transition to
a metallic spin glass; the local moment formation is therefore not directly
related to the metal-insulator transition.
We also review the the mean-field theory of the disordered Fermi liquid
to metallic spin glass transition and describe the separate
regime of ``non-Fermi liquid'' behavior
at higher temperatures near the quantum critical
point.
The relationship to experimental results on doped semiconductors and
heavy-fermion compounds is noted.
\end{abstract}

\section{Introduction}
This paper deals with the rich variety of magnetic phenomena
and phases that appear in strongly disordered and correlated
electronic systems in the vicinity of a metal-insulator
transition. In particular, much attention has been focussed on
the ubiquitous ``local moments'' which appear to dominate
the low temperature thermodynamics of the disordered metallic
state, and also across the metal-insulator transition into the
paramagnetic insulator (Quirt \& Marko 1971; Ue \& Maekawa 1971; Alloul \&
Dellouve 1987;
Sachdev 1989; Milovanovic {\it et al.} 1989; Bhatt \& Fisher 1992;
Tusch \& Logan 1993, 1995; Lakner {\it et al.} 1994; Langenfeld \& Wolfle
1995).
The principal thesis of this article is that these local moments are
{\it not} directly related to the critical degrees of freedom leading to the
metal-insulator transition. Rather, they should be understood as the ``quantum
Griffiths''
precursors of the transition from a disordered metal to a {\it metallic spin
glass}.

Below, we will review the basic ideas needed to motivate and explain this
thesis.
We will also discuss a recent mean field theory of the transition from the
metal
to the metallic spin glass (Sachdev {\it et al.} 1995; Sengupta \& Georges
1995)
and review its experimental consequences. We will not discuss the
metal-insulator
transition in the main part of the paper, but will speculate on the
implications of
our ideas for it in the concluding section.

\section{General considerations at zero temperature}
\label{general}
We will phrase our discussion in terms of the following Hamiltonian
\begin{equation}
{\cal H} = -\sum_{i < j , \alpha} t_{ij} c_{i\alpha}^{\dagger} c_{i\alpha}
+ U \sum_i n_{i\uparrow} n_{i\downarrow} -\sum_{i,\alpha}
(\epsilon_i - \mu) c_{i \alpha}^{\dagger} c_{i \alpha}
\label{hamgeneral}
\end{equation}
where $c_{i\alpha}$ annihilates an electron on site $i$ with spin
$\alpha=\uparrow, \downarrow$, and $n_{i\alpha} = c^{\dagger}_{i\alpha}
c_{i\alpha}$.
The sites are placed in three dimensional space and labeled by
$i,j$. The electrons are in a chemical potential $\mu$ and repel each other
with the on-site repulsion energy $U$. The hopping matrix elements between
the sites are the $t_{ij}$ which are short-ranged and possibly random.
Finally, we also allow for some randomness in the on-site energies,
$\epsilon_i$.
A Hamiltonian of the form of ${\cal H}$ is expected to be a good qualitative,
if not
quantitative, model of electronic motion in the impurity band
of doped semiconductors, and this fact is the primary reason for focussing on
it here. However, most of the discussion in this section is rather general
and should apply to a wide variety of strongly disordered electronic systems.

We begin with a simple but important question: on general grounds, what are the
different
possible ground states of ${\cal H}$ between which it is possible to make a
sharp distinction ?
Here we are considering two states distinct if a thermodynamic phase transition
is
required to connect one to the other. Notice also that we are referring to
phases at {\it zero temperature} ($T$) so the phase transitions are necessarily
quantum in
nature. In some cases, two phases separated by a quantum phase transition
at zero temperature can be connected smoothly at any nonzero temperature, and
so
the sharp distinctions made below are special to $T=0$.

\begin{figure}[tb]
\epsfxsize=4in
\centerline{\epsffile{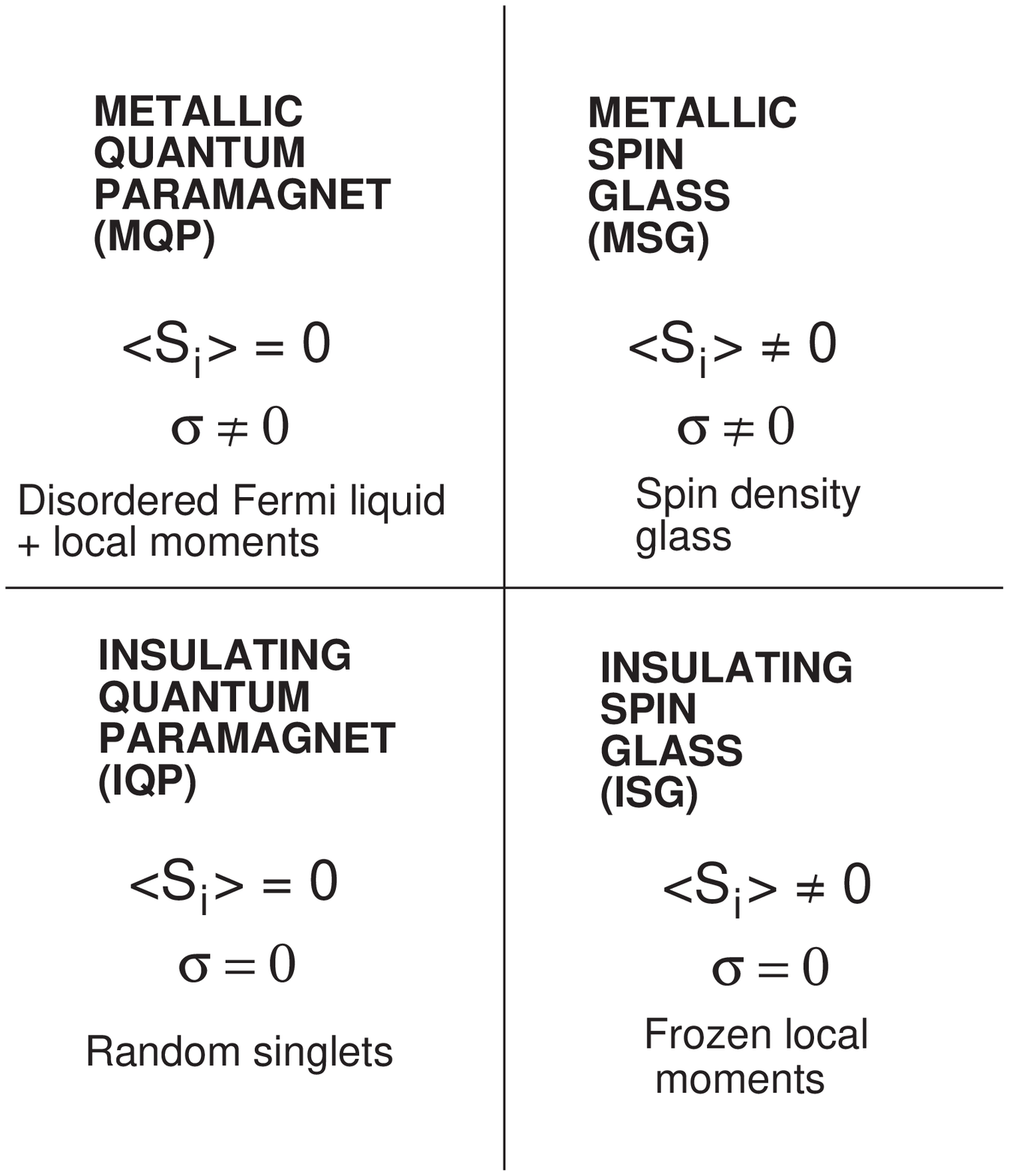}}
\vspace{0.2in}
\caption{Schematic diagram of the $T=0$ phases of a strongly random
electronic system in three dimensions described {\em e.g.\/}
by the Hamiltonian (\protect\ref{hamgeneral}). The average moment,
$\langle S_i \rangle$, when non-zero, varies randomly from site
to site. The conductivity is denoted by $\sigma$.
The phase diagram is a section in the parameter space of
$U/\mbox{(mean value of the $t_{ij}$)}$,
the width of the distribution of the $t_{ij}$,
the range of the $t_{ij}$, and the filling fraction.
}
\label{fig1}
\end{figure}

Figure~\ref{fig1} shows a schematic $T=0$ phase diagram of ${\cal H}$ as its
couplings are
varied (an explicit computation of a related phase diagram on a cubic lattice
has
been presented by Tusch \& Logan (1993,1995)).
The phase diagram is a section in the parameter space of
$U/\bar{t}$ (where $\bar{t}$ is the mean value of the $t_{ij}$),
the width of the distribution of the $t_{ij}$,
the range of the $t_{ij}$, and the filling fraction.
The phases are distinguished by the behavior of their
spin and charge fluctuations. Charge transport is characterized by the $T=0$
value of the conductivity, $\sigma$; if $\sigma = 0$ the phase is an insulator,
and is metallic otherwise. For the spin sector, the uniform spin susceptibility
is not a useful diagnostic (as will become clear below),
and we distinguish phases by whether the ground state has infinite memory of
the spin orientation on a site or not. The time-averaged moment on a given site
is denoted by $\langle \vec{S}_i \rangle$ (where
\begin{equation}
\vec{S}_i = c_{i\alpha}^{\dagger} \vec{\sigma}_{\alpha\beta} c_{j \beta}
\end{equation}
with $\vec{\sigma}$ the Pauli matrices) and it can either vanish at every site,
or take a non-zero value which varies randomly from site to site.
Experimentally, a phase with $\langle \vec{S}_i \rangle $ non-zero will
have an elastic delta function at zero frequency in the neutron
scattering cross section.
The average
over all sites will be denoted by $\overline{\langle \vec{S}_i \rangle}$,
and will be
non-zero only in ferromagnetic phases, which we will not consider here.

The four phases in Figure~\ref{fig1} are:

\vspace{0.1in}\noindent
(1) METALLIC QUANTUM PARAMAGNET (MQP)
\newline
More simply known as the familiar `metal', this phase has
\begin{equation}
\sigma \neq 0 ~~~~~~~~~~\langle \vec{S}_i \rangle = 0.
\end{equation}
The standard picture of this phase is in terms of low
energy excitations consisting of spin-1/2, charge $e$, fermionic,
itinerant quasiparticles.
The quasiparticles obey a transport equation as in
Landau's Fermi liquid theory, but have wave functions which are spatially
disordered.
Many low temperature transport properties of these quasiparticles have been
computed in a weak disorder expansion (Altshuler \& Aronov 1983;
Finkelstein 1983, 1984; Castellani \& DiCastro 1985; Belitz \& Kirkpatrick
1994).
Here, in Section~\ref{lm}, we will
review more recent work (Milovanovic {\it et al.} 1989; Bhatt \& Fisher 1992;
Lakner {\em et al.} 1994; Langenfeld \& Wolfle 1995)
that argues that this disordered Fermi liquid
is in fact unstable towards the formation of `local moments' of spin which
modify most of its thermodynamic properties. A careful definition of
a local moment, and its physical properties will appear
in Section~\ref{lm}, but loosely speaking, a local moment is a site
with a strongly localized charge $e$ and relatively slow spin fluctuations.
The spin fluctuations, although long-lived, eventually lose memory of their
orientation, and we always have $\langle \vec{S}_i \rangle = 0$.
There are large fluctuations in the lifetime of the spin orientations on the
local
moment sites, and those with the longest lifetimes may be considered as
the ``quantum Griffiths'' precursors (defined more carefully later)
of the metallic spin glass to be
considered below.

\vspace{0.1in}\noindent
(2) INSULATING QUANTUM PARAMAGNET (IQP)
\newline
This phase has
\begin{equation}
\sigma = 0 ~~~~~~~~~~\langle \vec{S}_i \rangle = 0
\end{equation}
and is accessed by a metal-insulator transition from the MQP phase.
The itinerant quasiparticles have now localized: this prevents them
from contributing to a d.c. conductivity, but many of their thermodynamic
effects are expected to be similar to those in the MQP phase, particularly
when the quasiparticle localization length is large. There is however no
sharp distinction between the localized quasiparticles and the local moments,
and at scales larger than the localization length one may view the IQP
simply as a collection of randomly located spins with $S=1/2$ and interacting
with one another by some effective exchange interaction.
There is a  large density of low energy spin excitations, which means that the
spin
susceptibility is quite large, and may even diverge as $T \rightarrow 0$ (Ma
{\it et al.}
1979; Dasgupta \& Ma 1980; Hirsch 1980; Bhatt \& Lee 1982).
This is the reason we
have not used the spin susceptibility as a diagnostic for the phases.
An exact solution of an IQP model with infinite-range exchange was presented
recently by Sachdev \& Ye (1993).

\vspace{0.1in}\noindent
(3) METALLIC SPIN GLASS (MSG)
\newline
This phase has
\begin{equation}
\sigma \neq 0 ~~~~~~~~~~\langle \vec{S}_i \rangle \neq 0,
\end{equation}
and is accessed from the MQP by a spin-freezing transition. The local moments
of the MQP phase now acquire a definite orientation, and retain memory of
this
orientation for infinite time. The Fermi liquid quasiparticle excitations are
still present and are responsible for the non-zero $\sigma$; the frozen
moments appear as random local magnetic fields to the itinerant quasiparticles.
Alternatively, we may view the spin-freezing transition as the onset, from
the MQP phase, of a spin density wave with random offsets in its phase and
orientation, as appears to be the case in recent experiments (Lamelas {\it et
al.}
1995); this
suggests the name ``\underline{spin density glass}''.
The spin density glass point of view was explored by
Hertz (1979) some time ago, but he did not focus on the vicinity of the
$T=0$ transition between the MSG and MQP phases.
Here, in Section~\ref{msg} we will review the recent complete solution for the
MSG phase and the MSG-MQP transition in the infinite range model (Sachdev {\it
et al.}
1995; Sengupta \& Georges 1995),
aspects of the Landau theory for the short range case
(Read {\it et al.} 1995; Sachdev {\it et al.} 1995), and recent experimental
realizations.
A more detailed review of this phase may be found elsewhere (Sachdev \& Read
1996).

\vspace{0.1in}\noindent
(4) INSULATING SPIN GLASS (ISG)
\newline
This phase has
\begin{equation}
\sigma = 0 ~~~~~~~~~~\langle \vec{S}_i \rangle \neq 0
\end{equation}
Charge fluctuations are unimportant, and the collective frozen spin
configuration is
expected to be well described by an effective classical spin model. Examples of
phases of this type may be found in well known reviews (Binder \& Young 1986;
Fischer \& Hertz 1991). We will not have anything
to say about this phase here.

Let us emphasize that the above discussion was restricted to $T=0$, and thus
describes only the ground state properties of $H$.
We will discuss the finite $T$ properties below. Well away from any of the
quantum phase %%@
boundaries, the characteristics
of the ground state are usually enough to give us a physical picture of the
low $T$ properties. However, closer to the quantum phase boundaries or at
higher
$T$, entirely new regimes emerge which cannot be understood in terms of
any of the phases describe above; rather, they are characterized by
the critical states at the phase boundaries, and their universal
response to temperature (Sachdev \& Ye 1992; Sachdev 1995).
This aspect of the physics will be explored
in our discussion of the MQP-MSG quantum phase transition.

\section{Formation of local moments in disordered metals}
\label{lm}
This section is about the magnetic properties of the MQP phase.
The popular ``weakly-disordered Fermi liquid'' approach to this phase
computes its low temperature properties by an expansion in the strength of the
random
elastic scattering by impurities (Altshuler \& Aronov 1983;
Finkelstein 1983, 1984; Castellani \& DiCastro 1985; Belitz \& Kirkpatrick
1994).
The main purpose of this section
is to show that this weak scattering expansion misses an important piece of
physics that actually dominates the low temperature thermodynamics and spin
transport.
The key phenomenon is the local instability of the interacting, disordered,
Fermi liquid to the appearance of magnetic moments whose slow dynamics controls
the long-time spin correlations. An essential property of this instability is
that it is caused by fluctuations in the disorder strength over
a small, localized region of space; it is not related to any development of
coherence
over large spatial scales. It may be viewed as another realization of the
general
class of ``quantum Griffiths'' effects which have been very important in
numerous
recent studies of quantum phases and phase transitions in random spin systems
(Fisher 1992, 1995; Thill \& Huse 1995; Read {\it et al.} 1995;
Guo {\it et al.} 1996;
Senthil \& Sachdev 1996).
In the present case these Griffiths effects are argued to be precursors of the
metallic
spin glass phase. That they disrupt what is usually considered to be an
analysis of the
metal-insulator transition is accidental: we argue they are not directly
related
to the metal-insulator transition and have clouded a proper interpretation of
theories
of it.
We will begin by defining more carefully what we mean by a local moment
in Section~\ref{lm1}. Then, in Section~\ref{lm2},
we will describe the current theoretical understanding of
the formation of these local moments in random systems, and their implications
for experiments.

\subsection{What is a local moment ?}
\label{lm1}
Unlike the sharply distinguishable zero temperature
phases that were discussed in Section~\ref{general},
the concept of a local moment initially appears somewhat imprecise,
and refers to intermediate energy phenomena at moderate, nonzero temperatures.
Alternatively, we may take a renormalization group point of view,
and introduce local moments, as one renormalizes down from higher energies,
as the degrees of freedom
necessary for a proper description of the low energy excitations about the
ground state.
Despite this seeming impreciseness,
as we shall see below, the presence of local moments does
lead to a qualitatively different description of the low temperature
properties of disordered metals.

\begin{figure}[tb]
\epsfxsize=4in
\centerline{\epsffile{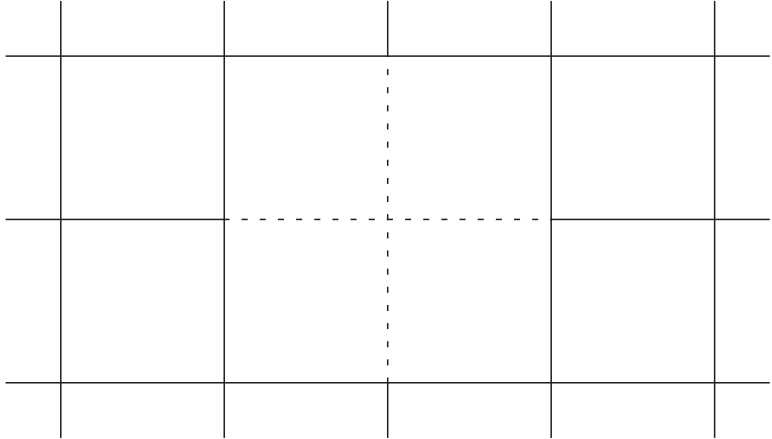}}
\vspace{0.2in}
\caption{Hopping matrix elements of the single-impurity model considered
in Section~\protect\ref{lm1}.
The dashed lines about the origin represent bonds with
$t_{ij} = w$, while all remaining
bonds  have $t_{ij} = t$.
}
\label{fig2}
\end{figure}

To explain the concept further, it is useful to specialize to a particular
realization of ${\cal H}$ (Milovanovic {\it et al.} 1989):
a single impurity model related to models considered
early on by Anderson (1961) and Wolff (1961).
Place all the sites, $i$, on the vertices of a regular lattice,
and set $\epsilon_i = 0$, $t_{ij} = t$ between all nearest neighbors, and
$t_{ij} = 0$
otherwise. Now pick a special `impurity' site $i=0$, and set all its nearest
neighbor bonds $t_{0i} = w$. (See Fig~\ref{fig2}).
.
We choose the chemical
potential $\mu = -U/2$ so that the system is half-filled, and the resulting
model
is now characterized by two dimensionless parameters: $w/t$ and $U/t$.
We will now argue, using the results of extensive numerical and analytic
studies on closely related models (Haldane 1978; Krishnamurthy {\it et al.}
1980),
that there are two qualitatively distinct regimes in the $w/t$ and $U/t$ plane,
characterized by very different behavior in their nonzero temperature
properties.

We will characterize the two regions by their local spin susceptibility,
$\chi_L$. This is
their response to a magnetic field coupled to the site $0$ under which
\begin{equation}
{\cal H} \rightarrow {\cal H} - \frac{g \mu_B H}{2}\left(
c_{0\uparrow}^{\dagger}
c_{0 \uparrow} - c_{0\downarrow}^{\dagger} c_{0 \downarrow} \right)
\end{equation}
First consider the very high temperature limit $T \gg t,U,w$. In this case all
states
are equally probably. On site $0$ there are 4 possible configurations: empty,
two
electrons, one spin up electron, and one spin down electron. Of these, the
latter
two contribute a standard Curie susceptibility of $(g \mu_B )^2 / ( 4 k_B T)$,
while
the first two have zero susceptibility.
The average high temperature susceptibility is therefore
\begin{equation}
\chi_L = \frac{(g \mu_B )^2}{8 k_B T}~~~~~~~T \gg U, t, w
\label{hight}
\end{equation}

Now lower the temperature. The implication of the earlier work (Anderson 1961;
Wolff 1961; Haldane 1978; Krishnamurthy {\it et al.} 1980) is that there
are two distinct possibilities in the $w/t$, $U/t$ phase diagram:

\vspace{0.1in}\noindent
(I) {\it Itinerant quasiparticle regime}
\newline
The value of $T \chi_L$ decreases monotonically from the high $T$ limit
(\ref{hight}) as $T$ is lowered. The states at the site $0$ are absorbed into
the itinerant quasiparticle states of the surrounding sites, and the
excitations
at $0$ are typical of those of a Fermi liquid. As $T \rightarrow 0$ we
therefore
expect the constant Pauli susceptibility $\chi_L \sim (g \mu_B)^2 / E_F$
where $E_F \sim t$ is the Fermi energy. Therefore $T \chi_L \rightarrow 0$
as $T \rightarrow 0$. This regime is clearly connected to the $w=t$ point where
the site $0$ ceases to be special.

\vspace{0.1in}\noindent
(II) {\it Local moment regime}
\newline
Now as $T$ is lowered below $T \sim U$, the value of $T \chi_L$ rises and
reaches
a plateau where
\begin{equation}
T \chi_L = \frac{(g \mu_B)^2}{4 k_B T}~~~~~~~T_K \ll T \ll U
\label{lowt}
\end{equation}
(The lower limit is the Kondo temperature $T_K$ and will be discussed below.)
The natural interpretation of this plateau is that two of the four
possible states at site $0$ have been suppressed: the site $0$
either has one spin up electron, or one spin down electron,
and it is extremely unlikely that it is in the state with no electrons
or with two electrons in a spin singlet state.
When this happens, we assert that a local moment has formed at site 0.
Note that this strong local correlation between the spin up and spin down
occupations is not compatible with an itinerant fermionic quasiparticle
description, and really requires one to think in terms of a fluctuating
spin (a local moment) at site $0$.
As the temperature is lowered further in this regime, the nontrivial physics
of the Kondo effect becomes active at temperatures of order
$T_K \sim t \exp ( - c U t/w^2)$ where $c$ is a constant of order unity.
Notice that this temperature is exponentially small for large $U$,
and the temperature range in (\ref{lowt}) is well defined.
At temperatures below $T_K$ the local moment at 0 is quenched by the formation
of
a spin singlet
with an electron drawn from the conduction band. The local suscepbility
therefore
eventually returns to that of the Pauli susceptibility of a Fermi liquid.

\begin{figure}[tb]
\epsfxsize=4in
\centerline{\epsffile{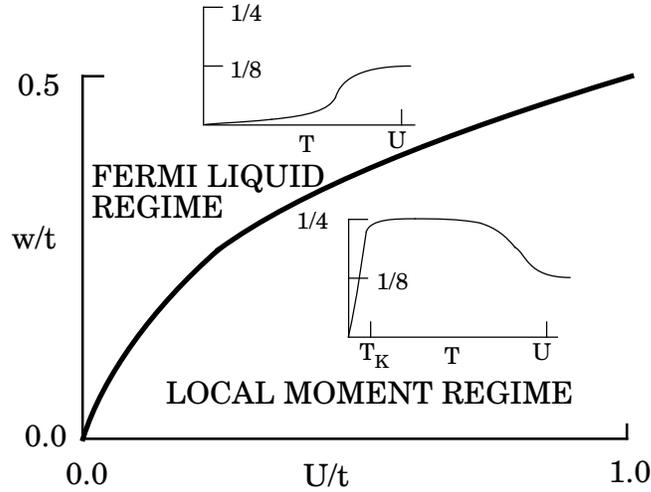}}
\vspace{0.2in}
\caption{Mean field phase diagram of the single impurity Hubbard model
sketched in Fig~\protect\ref{fig2}. The solid line represents a smooth
crossover which
appears as a sharp transition in the mean-field theory.
The insets plot the values of $T\chi_L$ in the
two regimes. (From Milovanovic {\it et al.} (1989))}
\label{fig3}
\end{figure}

Notice that as $T \rightarrow 0 $, there is no sharp distinction between cases
(I)
and (II) above: in both cases the spin susceptibility is that of a Fermi
liquid.
Nevertheless, there is a significant difference in the intermediate temperature
behavior. In the original mean field theory of Anderson (1961) this distinction
between the two cases appears as a sharp phase transition, but this is an
artifact of the mean-field approximation.
We show in Fig~\ref{fig3} the phase diagram of the single impurity model
as determined by the generalization of the Anderson (1961) mean field theory.
Also sketched
are the behaviors of $T \chi_L$ outlined above.

It is also useful to notice the analogy between the features noted above and
the fluctuations which lead to the quantum Griffiths singularities in random
spin systems (Fisher 1992, 1995; Thill \& Huse 1995; Read {\it et al.} 1995;
Guo {\it et al.} 1996; Senthil \& Sachdev 1996). In the latter cases the
important random %%@
fluctuations
are in localized regions in which the quantum spin correlations have their
values
enhanced by some finite fraction. As above, the contribution of any single such
region is eventually quenched by its quantum coupling to the surrounding spins.
The true importance of the local random fluctuations, and the qualitative
difference
they make to the macroscopic properties, only becomes apparent when the
collective statistics of the global system is examined. We will turn to this in
the
following subsection.

\subsection{Instabilities in the fully disordered model}
\label{lm2}
Now we turn to the full complexity of ${\cal H}$, allowing the configuration of
every
site to be random. We anticipate that quantum Griffiths effects noted above
could have an important influence here, but before we can apply such arguments,
more elementary issues need to be settled:
\newline
({\em i}) Do the single impurity
instabilities discussed above apply to the fully random case, in which every
site is potentially an impurity site ?
\newline
({\em ii}) What is the spatial extent of each local moment instability ? In
particular,
is it correlated with the localization length of the itinerant quasiparticles ?
\newline
({\em iii}) Is it valid to treat the different instabilities independently of
each other ?
\newline
Although the possibile importance of local moments to disordered metals had
been noted earlier (Quirt \& Marko 1971; Ue \& Maekawa 1971; Alloul \& Dellouve
1987;
Sachdev 1989), these questions were first addressed by Milovanovic {\em et al.}
(1989), and we
review their results below.

First, notice that the distinction between the two regimes of the single
impurity model
became apparent at a relatively high temperature $T \sim U$. The same feature
is expected
to hold in the fully disordered model, and we will therefore treat the
interactions
in ${\cal H}$ within the Hartree-Fock approximation. It is essential, however,
to include
the effects of disorder exactly: this approach is therefore the converse of the
method used to describe the disordered itinerant quasiparticles (Altshuler \&
Aronov 1983;
Finkelstein 1983, 1984; Castellani \& DiCastro 1985; Belitz \& Kirkpatrick
1994).
A useful way to formulate the Hartree-Fock or `self-consistent field' approach
is to
consider it as the optimization of a single-particle effective Hamiltonian
${\cal H}_{\rm eff}$.
We choose this in the form
\begin{equation}
{\cal H} = -\sum_{i < j , \alpha} t_{ij} c_{i\alpha}^{\dagger} c_{i\alpha}
+ \sum_i (\widetilde{\epsilon}_i - \mu) c_{i \alpha}^{\dagger} c_{i \alpha}
- \sum_i \vec{h}_i \cdot \vec{S}_i
\end{equation}
where $\widetilde{\epsilon}_i$ and $\vec{h}_i$ are the variational parameters.
Notice that the $\vec{h}_i$ are local magnetic fields which polarize the
electrons
on site $i$. The appearance of a significant $\vec{h}_i$ on a given site is
the signal of a formation of a local moment, and the spatial form
of the $\vec{h}_i$ field therefore contains the essential information
that we want. As we are mainly discussing the issue
of the initial instability towards the formation of local moments, it is useful
to
expand the effective free energy in powers of the $\vec{h}_i$. The results of
such an
expansion are
\begin{equation}
{\cal F}_{\rm eff} ( \vec{h}_i ) = {\cal F}_0 +
\frac{1}{4} \sum_{i,j,k} \chi_{ij} ( \delta_{jk} - U \chi_{jk} ) \vec{h}_i
\cdot
\vec{h}_k + {\cal O} ( \vec{h}^4 )
\label{freeen}
\end{equation}
where ${\cal F}_0$ is the free energy of the unpolarized state with $\vec{h}_i
= 0$.
The quantity $\chi_{ij}$ is the spin susceptibility matrix of this unpolarized
state---
it is the response in the magnetization at site $i$ to a magnetic field at site
$j$.
More specifically, it is defined by
\begin{equation}
\chi_{ij} = \sum_{\alpha,\beta} \Psi_{\alpha} ( i) \Psi_{\beta}^{\ast} ( i)
\Psi_{\alpha}^{\ast} ( j )  \Psi_{\beta} (j) \frac{f(\lambda_{\alpha}) -
f(\lambda_{\beta})}{
\lambda_{\beta} - \lambda_{\alpha}}
\label{defchi}
\end{equation}
where $f(\lambda)$ is the Fermi function,
$\Psi_{\alpha} (i)$ are the wavefunctions of the itinerant quasiparticles in
the
unpolarized state, and $\lambda_{\alpha}$ are their energies. In the present
effective
field approximation these are the eigenvalues and eigenenergies of a
one-particle
Hamiltonian
\begin{equation}
\sum_{j} \left[ (\widetilde{\epsilon}_i - \mu) \delta_{ij} -t_{ij} \right]
\Psi_{\alpha} (j) =
\lambda_{\alpha} \Psi_{\alpha} ( i) .
\label{psi}
\end{equation}
Finally, the parameters $\widetilde{\epsilon}_i$ are determined by the
nonlinear
self-consistency condition
\begin{equation}
\widetilde{\epsilon}_i = \epsilon_i + U \sum_{\alpha} \left|
\Psi_{\alpha} (i) \right|^2 f ( \lambda_{\alpha} ) .
\label{epsilon}
\end{equation}
The equations (\ref{psi}) and (\ref{epsilon}) were solved numerically for
realistic realizations
of the disorder appropriate to $Si:P$. The density of the disorder was chosen
so that the system is comfortably within the metallic phase. The quasiparticle
wavefunctions, %%@
$\Psi_{\alpha}$, are
extended for a wide range of energies near the Fermi level, and
explicit evidence is presented for this below.
{}From the known $\Psi_{\alpha}$ and $\lambda_{\alpha}$, the susceptibility
matrix $\chi_{ij}$
was determined from (\ref{defchi}).

To proceed further, is is useful to now diagonalize $\chi_{ij}$. We solve the
eigenvalue
problem
\begin{equation}
\chi_{ij} m_{a} (j) = \kappa_a m_a (j)
\end{equation}
where $\kappa_a$ are the eigenvalues, and $m_a$ are the normalized
eigenvectors. Now make
the expansion
\begin{equation}
\vec{h}_i = \sum_a \vec{p}_a m_a ( i)
\end{equation}
and insert it into (\ref{freeen}). We get
\begin{equation}
{\cal F} = {\cal F}_0 + \frac{1}{4} \sum_a \kappa_a ( 1 - U \kappa_a )
\vec{p}_a^2
+ {\cal O} (\vec{p}^4)
\end{equation}
We see that there is first an instability to a non-zero value of $\vec{p}_{a}$
if
the associated eigenvalue $\kappa_a > 1/U$. We identify the temperature at
which this
first happens with the formation of a local moment with spatial distribution of
spin
proportional
to $m_a ( i)$. What now happens at lower temperatures depends upon the spatial
structure of
$m_a (i)$. If all of the instabilities appear in {\em localized} eigenvectors
$m_a (i)$
which are spatially well separated from each other, then they are approximately
decoupled, and we can identify the formation of each of those local moments
at the point where their respective $\kappa_a > 1/U$. On the other hand, if any
of the $m_a (i)$ are extended, then we have to go into the magnetized phase
with $\vec{p}_a = 0$, recompute the new susceptibility matrix, and diagonalize
it again. The latter possibility in fact corresponds to the appearance of the
long-range
order of the metallic spin glass phase, and will be discussed in
Section~\ref{msg}.
In the present circumstances we found that all the $m_a$ were indeed
well localized.

\begin{figure}[tb]
\epsfxsize=4.5in
\centerline{\epsffile{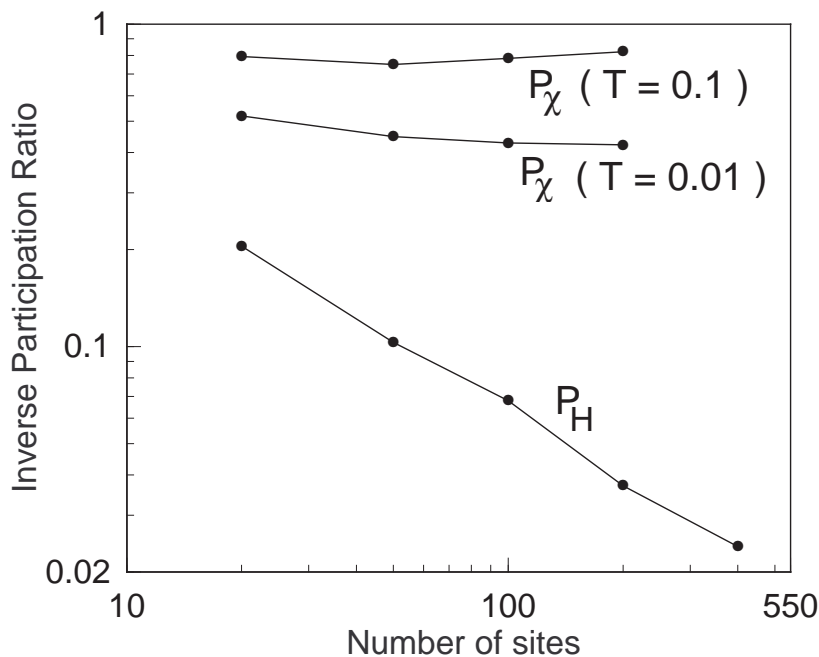}}
\vspace{0.15in}
\caption{Evidence for the localization of the magnetization eigenvectors in the
metallic phase. The inverse participation ratios of the eigenvectors of the
susceptibility ($P_{\chi}$) and the quasiparticle Hamiltonian ($P_{{\cal H}}$)
plotted as a function of system size. Notice that the magnetic modes are
well localized as $P_{\chi}$ is roughly independent of system size,
while the quasiparticle excitations are extended as $P_{{\cal H}}$ decreases
rapidly
with system size.(From Milovanovic {\it et al.} (1989))}
\label{fig4}
\end{figure}

Evidence for the localization is presented in Fig~\ref{fig4}.
There we compare the
values of the inverse participation ratios $P_{{\cal H}}$ and $P_{\chi}$ which
are defined
by
\begin{equation}
P_{{\cal H}} = \left\langle \sum_i |\Psi_{\alpha} ( i) |^4 \right\rangle~~~~~~~
P_{\chi} = \left\langle \sum_i |m_a ( i) |^4 \right\rangle
\end{equation}
The eigenstates $\Psi_{\alpha} ( i) $ and $m_a (i)$ are both normalized,
and so $P_{{\cal H},\chi}$ are measures of their average spatial extent. If the
states are localized then $P_{{\cal H},\chi}$ will saturate to some finite
value
of order unity as the system size is increased; on the other hand, if the
states
are extended, we expect $P_{{\cal H},\chi}$ to roughly decrease as $1/N$ where
$N$ is
the number of sites. We see from Fig~\ref{fig4} that the behavior of the two
participation
ratios is dramatically different. The quasiparticle wavefunctions
$\Psi_{\alpha} (i)$
are clearly extended, as has been previously claimed. For the same samples,
however,
the $m_a (i)$ are well localized, on the scale of just one or two sites
($P_{\chi}
\sim 0.5$).

The localization of the $m_a ( i ) $ is the key result of the numerical
analysis of Milovanovic {\em et al.} (1989).
It answers the questions posed at the beginning of this subsection.
This localization justifies the identification of the instabilities
in the fully random ${\cal H}$ and ${\cal H}_{\rm eff}$ with essentially
independent
instabilities, much like those in the single impurity model. The local
moments are confined to just a few sites, and this localization is quite
independent of the localization length of the quasiparticles, which are in fact
itinerant. The initial instabilities at different points in the sample are
essentially decoupled as they correspond to different localized eigenvectors
of $\chi_{ij}$ with little mutual overlap.
All of these features are again consistent with their interpretation
as Griffiths effects which require rare, {\em independent\/}, fluctuations in
different
regions of the sample.
All Griffiths effects are precursors of some actual phase transition,
the proper identification of which requires some understanding of lower
temperature properties of the local moments, which we address in the next
section.

\subsection{Low temperature thermodynamics of local moments}
\label{lttlm}
Once the local moments have formed at temperatures $T \sim U$, a different
effective Hamiltonian for the low temperature properties of the metal
is clearly warranted. On the local moment sites, states with two spin singlet
electrons, or with no electrons, occur only with a small probability,
and it will clearly pay to eliminate them by a canonical transformation.
Let us denote that sites upon which we want to perform this canonical
transformation by $\ell$. Then on each such site $\ell$ we have only two
allowed
states: a spin up electron or a spin down electron, and a quantum spin-$1/2$
operator
$\vec{S}_{\ell}$ which transforms among them.

We have assumed above that the local moment resides on a single site.
It is clearly possible to have moments located on two or more sites,
but we expect that a similar procedure can be applied. A higher spin-$j$
operator might be required, and more states may have to be eliminated,
but these complications are not expected to modify any of the following
discussion.

We can now write down an effective
Hamiltonian for the entire random system
\begin{eqnarray}
{\cal H}_K = && -\sum_{i < j , \alpha}\! ' t_{ij} c_{i\alpha}^{\dagger}
c_{i\alpha}
+ U \sum_i\! '  n_{i\uparrow} n_{i\downarrow} -\sum_{i,\alpha}\! '
(\epsilon_i - \mu) c_{i \alpha}^{\dagger} c_{i \alpha} \nonumber \\
&&~~~~~~~~~~~~+ \sum_i\! '  \sum_{\ell} J_{i\ell} \vec{S}_{\ell} \cdot c_{i
\alpha}^{\dagger}
\vec{\sigma}_{\alpha\beta} c_{i \beta}.
\end{eqnarray}
The primes on the sums indicate that the sites $\ell$ have to be
excluded from the sum over $i$. The new feature is the Kondo exchange coupling,
$J_{i\ell}$
(expected to be antiferromagnetic) between the local moments and the itinerant
electrons, and this is the remnant of the canonical transformation eliminating
the states on the local moment sites.
The low temperature properties of random models like ${\cal H}_K$ have been
considered by a number of investigators
(Bhatt \& Fisher 1992;
Dobrosavljevic {\em et al.} 1992; Dobrosavljevic \& Kotliar 1993;
Tusch \& Logan 1993, 1995;
Lakner {\em et al.} 1994; Langenfeld \& Wolfle 1995; Rowan {\it et al.} 1995;
Miranda {\em et al.} 1996,1997)

We will focus here on the elegant results of Lakner {\em et al.} (1994) and
Langenfeld \& Wolfle (1995) which have the virtue of containing
{\em quantitative} predictions specific to $Si:P$. They began with a
quantitative
computation of the boundary between the single impurity local moment and
itinerant quasiparticle regimes in environments characteristic of those found
in $Si:P$. Assuming that the location of the $P$ sites in $Si:P$ were
statistically uncorrelated, they computed the probability distributions
of the local environments of the sites. The combination of these results
allowed to predict two quantities essential for a quantitative comparison
with experimental results ({\em i}) the density of local moments as a function
of $P$ doping, and ({\em ii\/}) the probability distribution of the Kondo
temperatures ${\cal P} ( T_K )$ of these local moment sites.
Over a broad range of $T_K$ values, including the entire experimentally
relevant range,
they found that
\begin{equation}
{\cal P} ( T_K ) = \frac{c_0}{T_{K}^{\alpha_K}}
\end{equation}
where $c_0$ is a normalization constant, and $\alpha_K \approx 0.9$.
If one now ignores the possible exchange couplings between the local moments
(these could be mediated by the itinerant quasiparticles, or due to a direct
exchange), the local moment contribution to
thermodynamic properties can be computed by a simple superposition of the
universal
single impurity thermodynamics at different Kondo temperatures. Such
a procedure gives a specific heat $C \sim T^{1- \alpha_K}$ and a
uniform spin susceptibility $\chi \sim T^{-\alpha_K}$. Lakner {\em et al.}
(1994)
find good quantitative agreement between such model calculations
and experimental measurements of the specific heat.

The singular low temperature effects of the local moments above are
clearly associated with those sites at which $T_K$ is anomalously small.
The local spin orientation on those sites has a long lifetime,
and under these conditions any RKKY coupling between them will clearly be
important,
although it has been neglected in the above analyses.
Indeed this RKKY coupling will prefer a fixed relative orientation among
the slowly fluctuating moments. As the RKKY coupling is an oscillating function
of separation, the relative orientation will oscillate in sign.
Taking the limit over which this RKKY coupling
correlates an increasingly larger number of local moments, we see that
eventually
spin glass order will appear. None of this incipient ordering has directly
involved
the itinerant quasiparticles, and so the system will remain metallic.
These arguments clearly show that the Griffiths effects associated with
local moment formation are precursors of the metallic spin glass phase.

\section{Metallic spin glasses}
\label{msg}

We now turn to an analysis of the metallic spin glass (MSG), with a particular
focus on its transition to the metallic phase (MQP) and the crossovers
at finite temperature in the vicinity of the quantum critical point.
This analysis will be done within the framework of a Landau theory
for this transition that was developed recently (Read {\it et al.} 1995;
Sachdev {\it et al.} 1995). Such analysis effectively focuses only
on the extended eigenmodes of the susceptibility $\chi_{ij}$ into which
there is condensation in the spin glass phase. The quantum Griffiths effects
that were crucial to understanding the low temperature properties of the
MQP are entirely missing at the mean field level. Omission of these
Griffiths effects is probably quite dangerous at low temperatures. However, we
shall find new regimes at moderate temperatures which display novel physics
associated
with the extended states of the critical point, and the analysis below should
be considered a first attempt towards their understanding.

Portions of the discussion below are adapted from the recent review article by
Sachdev \& Read (1996), and the reader is refered to it for greater detail.
We shall take a Landau theory point of view in this article, and hence
automatically
obtain a formalism which is suited for analysis of fluctuations (Sachdev {\it
et al.}
1995; Sachdev \& Read 1996). However it is also possibly to obtain results
equivalent
to those of the Landau mean field theory by the exact solution of an
infinite-range
model: such an approach was followed by Sengupta \& Georges (1995) (and also
for related models by Huse \& Miller (1993) and Ye {\it et al.} (1993)),
but we shall not use it here.

In the following subsections
we will ({\em i\/}) introduce the order parameter for quantum spin glass
to paramagnet transition, ({\em ii\/}) obtain a
Landau functional for this order parameter,
({\em iii\/}) minimize the functional as a function of temperature
and coupling constant to obtain the predictions of crossover functions,
and ({\em iv\/}) briefly discuss the relationship to recent experiments on
heavy fermion compounds.

\subsection{Order parameter}
We begin by introducing the order parameter for the quantum phase
transition (Read {\it et al.} 1995).
Recall that for classical spin glasses in the replica formalism, this is a
matrix $q^{ab}$,
$a,b =1 \ldots n$ are replica indices and $n \rightarrow 0$.
The \underline{off-diagonal} components of $q^{ab}$ can be related to the
Edwards-Anderson order parameter, $q_{EA}$, in a somewhat subtle way we won't
go into
here (Fischer \& Hertz 1991; Binder \& Young 1986).
In quantum ($T=0$) phase transitions, time dependent
fluctuations of the order parameter must be considered (in ``imaginary''
Matsubara time $\tau $), and in the spin glass case it is found that the
standard
decoupling, analogous to the classical case introducing $q^{ab}$, leads
now to a matrix function of two times (Bray \& Moore 1980)
which we can consider to be
\begin{equation}
Q^{ab} ( x , \tau_1 , \tau_2 ) =  \sum_{i \in {\cal N}(x)}
S_{i}^a ( \tau_1 )
S_{i}^b (\tau_2 )
\label{composite}
\end{equation}
where ${\cal N}(x)$ is a coarse-graining region in the neighborhood of $x$,
and we will henceforth omit the vector spin indices on all operators except
where necessary.
{}From the set-up of the replica formalism it is clear that
\begin{eqnarray}
\overline{\langle S_i (0) \cdot S_i ( \tau ) \rangle} =
\lim_{n \rightarrow 0}
\frac{1}{n} \sum_{a} \left\langle\left\langle Q^{aa} ( x, \tau_1=0 ,
\tau_2=\tau) \right\rangle\right\rangle  \\
q_{EA} = \lim_{\tau \rightarrow  \infty} \lim_{n \rightarrow 0}
\frac{1}{n} \sum_{a} \left\langle\left\langle Q^{aa} ( x, \tau_1=0 ,
\tau_2=\tau) \right\rangle\right\rangle
\label{defqea}
\end{eqnarray}
relating $q_{EA}$ to the replica \underline{diagonal} components of $Q$.
We have introduced above double angular brackets to represent averages taken
with
the translationally invariant replica action (recall that single angular
brackets
represent thermal/quantum averages for a fixed realization of randomness,
and overlines represent averages over randomness).
Notice that the fluctuating field $Q$ is in general a function of two separate
times
$\tau_1$ and $\tau_2$; however the expectation value of its replica diagonal
components
can only be a function of the time difference $\tau_1 - \tau_2$. Further,
the expectation value of the replica
off-diagonal components of $Q$ is independent of both $\tau_1$ and $\tau_2$ (Ye
{\it et al.}
1993), and has a structure very similar to that of the classical order
parameter $q^{ab}$.
One can therefore also obtain $q_{EA}$ from the replica off-diagonal components
of $Q$,
as noted above for $q^{ab}$. Let us also note for completeness
that, unlike the quantum case, the replica diagonal components of the
classical order parameter $q^{ab}$ are
usually constrained to be unity, and contain no useful information.

The order parameter we shall use is $Q^{ab} (x, \tau_1, \tau_2)$, which is
a matrix in a replica space
and depends on the spatial co-ordinate $x$ and two times $\tau_1$, $\tau_2$.
However,
a little thought shows that this function contains too much information. The
important
degrees of freedom, for which one can hope to make general and universal
statements,
are the long-time spin correlations with $|\tau_1 - \tau_2| \gg \tau_m$,
where $\tau_m$ is a microscopic time like an inverse of a typical
exchange constant. As presented, the function $Q$ contains
information not only on the interesting long-time correlations,
but also on the uninteresting time range with $|\tau_1 - \tau_2|$
smaller than or of order $\tau_m$. The correlations in the latter range
are surely model-dependent and cannot be part of any general Landau action.
We shall separate out this uninteresting part of $Q$ by performing the shift
\begin{equation}
Q^{ab} (x, \tau_1 , \tau_2 ) \rightarrow
Q^{ab} (x, \tau_1 , \tau_2 ) - C \delta^{ab} \delta (\tau_1 - \tau_2)
\label{repar}
\end{equation}
where $C$ is a constant, and the delta function $\delta ( \tau_1 - \tau_2 )$
is a schematic for a function which decays rapidly to zero on a scale $\tau_m$.
The value of $C$ will be adjusted so that the resulting $Q$ contains only
the interesting long time physics: we will see later how this can be done
in a relatively straightforward manner.
The alert reader may recognize some similarity between the above procedure,
and the analysis of Fisher (1978) of the Yang-Lee edge problem.
In that case, too, the order parameter contains an uninteresting non-critical
piece which has to be shifted away; we will see below that there many other
similarities between the Yang-Lee edge and quantum spin glass problems.

\subsection{Action functional}
\label{action}
The action functional can be derived by explicit computations on
microscopic models or deduced directly from general arguments which
have been discussed in some detail by Read {\it et al.} (1995).
Apart from a single non-local term present in the metallic case (Sachdev {\it
et al.} (1995)
(see below), the remaining important terms are consistent with the general
criteria that:
\begin{enumerate}
\item The action is an integral over space of a local operator which can be
expanded in
gradients of powers of $Q$ evaluated at the same position $x$.
\item $Q$ is bilocal ({\em i.e.}\ is a matrix) in time, and each time is
associated with
one of the two replica indices (see
definition Eqn (\ref{composite})). These ``indices'' can appear more than once
in a term and
are summed over freely subject to the following rules before summations:
\begin{enumerate}
\item Each distinct replica index appears an even number of times.
\item Repetition of a time ``index'' corresponds to quantum-mechanical
interaction of spins, which must be
local in time and accordingly can be expanded as terms with times set
equal plus the same with
additional derivatives; it occurs when the corresponding replica
indices are the same, and only
then.
\end{enumerate}
\end{enumerate}

We now present all the terms, which, a subsequent renormalization-group
analysis tells
us are important near the quantum critical point. This is only a small subset
of
the terms allowed by the above criteria.

A crucial term is that linear in the order parameter $Q$. This term
encodes the local, on-site physics of the spin glass model, and tells
us the that spin is coupled to a metallic bath of itinerant quasiparticles.
\begin{equation}
\frac{1}{\kappa t} \int d^d x \left\{ \int d\tau
\sum_a \widetilde{r} Q^{aa} (x , \tau , \tau ) - \frac{1}{\pi} \int
d\tau_1 d\tau_2
\sum_{a} \frac{Q^{aa} (x, \tau_1, \tau_2)}{(\tau_1 - \tau_2 )^2}
\right\}
\label{actione2}
\end{equation}
The coupling
$\widetilde{r}$ will be seen below to be the critical tuning parameter for the
transition
from the spin glass to the paramagnet.
There is an overall factor of $1/\kappa t$
in front of this term; we have written this factor as a product of two coupling
constants, $\kappa$ and $t$, for technical reasons we shall not discuss
here.

There is a quadratic gradient term
\begin{equation}
\frac{1}{2t} \int  d^d x d \tau_1
d \tau_2 \sum_{a,b} \left[ \nabla Q^{ab} (x, \tau_1, \tau_2 )
\right]^2
\label{actione3}
\end{equation}
which is responsible for the development of spatial correlations in the
spin glass order. A quadratic term
without gradients
\begin{equation}
\int d^d x d \tau_1
d \tau_2 \sum_{a,b} \left[ Q_{\mu\nu}^{ab} (x, \tau_1, \tau_2 )
\right]^2
\label{actione4}
\end{equation}
is also allowed by the general criteria, but we choose to tune its coefficient
to
zero by using the freedom in (\ref{repar}). As will become clear in the next
subsection, this criterion is identical to requiring the absence of
uninteresting short time behavior in $Q$.
Notice again the formal similarity to the theory of the Yang-Lee edge (Fisher
1978),
where setting the coefficient of a quadratic term to zero was also responsible
for removing the uninteresting non-critical part of the order parameter
variable.

Next we consider cubic non-linearities, and the most important among the
several
allowed terms is the one with the maximum
number of different time and replica indices:
\begin{equation}
- \frac{\kappa}{3t} \int  d^d x d \tau_1 d \tau_2 d \tau_3
\sum_{a,b,c} Q^{ab} (x, \tau_1 , \tau_2 ) Q^{bc}
(x, \tau_2 , \tau_3 ) Q^{ca}
(x, \tau_3 , \tau_1 ).
\label{actione5}
\end{equation}
This term accounts for non-linearities induced solely by disorder
fluctuations.

Of the terms with fewer than the maximum allowed number of time indices
at a given order, the most important one is the one at quadratic order:
\begin{equation}
\frac{u}{2t} \int d^d x d \tau \sum_a  u~Q^{aa} ( x, \tau , \tau) Q^{aa}
( x, \tau , \tau).
\label{actione6}
\end{equation}
The coupling $u$ is the only one responsible for quantum mechanical
interactions between the spins, and as a consequence, all the time and
replica indices in (\ref{actione6}) are the same.

Lastly we have a final quadratic term
\begin{equation}
- \frac{1}{2t^2} \int d^d x \int  d \tau_1 d \tau_2
\sum_{a,b}
Q^{aa}  (x, \tau_1 , \tau_1 ) Q^{bb} ( x, \tau_2 , \tau_2 ),
\label{actione7}
\end{equation}
which accounts for the spatial fluctuation in the position of the
paramagnet-spin glass
transition. Recall that the linear coupling $\widetilde{r}$ was the control
parameter for this
transition, and a term like (\ref{actione7}) is obtained by allowing for
Gaussian
fluctuations in $\widetilde{r}$,
about its mean value, from point to point in space.
It will turn out that (\ref{actione7}) plays no role in the mean-field analysis
in the following subsection. However, it is essential to include
(\ref{actione7})
for a proper theory of the fluctuations.

The final Landau theory of the metallic spin glass and its transition to the
metal
is then (\ref{actione2})  + (\ref{actione3})  + (\ref{actione4})  +
(\ref{actione5})  +
(\ref{actione6})  + (\ref{actione7}).

\subsection{Mean field theory}
\label{mean}

We will now minimize the action of Section~\ref{action}.
We will review
the mean field theory for the MQP phase and identify the position of its
instability to the MSG phase. A discussion of the solution within the MSG phase
will not be presented here.

We Fourier transform from imaginary time to Matsubara frequencies by expressing
the action in terms of
\begin{equation}
Q^{ab} (x, \omega_1 , \omega_2 ) = \int_0^{1/T} d \tau_1 d \tau_2
Q^{ab} (x, \tau_1 , \tau_2 ) e^{- i ( \omega_1 \tau_1 + \omega_2 \tau_2 )},
\label{mft1}
\end{equation}
where we are using units in which $\hbar = k_B = 1$, and the frequencies,
$\omega_1$, $\omega_2$ are quantized in integer multiples of $2 \pi T$.
Then, we
make an ansatz for the mean-field value of $Q$ which is $x$-independent, and
dependent only on $\tau_1 - \tau_2$; within the MQP phase this takes the form
\begin{equation}
Q^{ab} (x, \omega_1, \omega_2 ) = (\delta^{ab} \delta_{\omega_1 +
\omega_2, 0}/T ) \chi_L (i \omega_1 )
\label{mft2}
\end{equation}
where we have used (\ref{defqea}) to identify the right hand side as the local
dynamic
susceptibility. Inserting (\ref{mft2}) into the action for the metallic case
in Section~\ref{action}, we get for the free energy per unit volume
${\cal F}/n$
(as usual, ${\cal F}/n$ represents the physical disorder averaged free energy):
\begin{equation}
\frac{{\cal F}}{n} = \frac{T}{t} \sum_{\omega} \left[
\frac{|\omega| + \widetilde{r}}{\kappa} \chi_L (i \omega )  - \frac{\kappa}{3}
\chi_L^3 (i \omega %%@
)
\right] + \frac{u}{2t} \left[ T \sum_{\omega} \chi_L ( i \omega ) \right]^2
\label{mft3}
\end{equation}
Notice that the coupling $1/t$ appears only as a prefactor in front of the
total
free energy, as the contribution of the $1/t^2$ term (\ref{actione7}) vanishes
in the replica
limit $n \rightarrow 0$. The value of $t$ will therefore play no role in the
mean field
theory. We now determine the saddle point of (\ref{mft3}) with respect to
variations
in the whole function $\chi_L ( i \omega )$, and find the solution
\begin{equation}
\chi_L ( i \omega ) = - \frac{1}{\kappa} \sqrt{ |\omega | + \Delta},
\label{mft4}
\end{equation}
where the energy scale $\Delta$ is determined by the solution of the equation
\begin{equation}
\Delta = \widetilde{r} - u T \sum_{\omega} \sqrt{|\omega | + \Delta }.
\label{mft5}
\end{equation}
Taking the imaginary part of the analytic continuation of (\ref{mft4})
to real frequencies, we get
\begin{equation}
\chi_L^{\prime\prime} ( \omega ) = \frac{1}{\kappa \sqrt{2}}
\frac{\omega}{\sqrt{\Delta + \sqrt{\omega^2 + \Delta^2}}}.
\label{cross2}
\end{equation}
Inserting the solution for $\chi_L$ back into (\ref{mft3}), and using
(\ref{mft5}), we get for the
free energy density:
\begin{equation}
\frac{{\cal F}}{n} = - \frac{1}{\kappa^2 t} \left[
\frac{2T}{3} \sum_{\omega} ( |\omega| + \Delta )^{3/2} + \frac{u}{2}
\left( T \sum_{\omega} \sqrt{ |\omega | + \Delta} \right)^2 \right]
\label{mft5a}
\end{equation}
The equations (\ref{mft5}), (\ref{cross2}), (\ref{mft5a}) are key
results (Sachdev {\it et al.} 1995; Sengupta \& Georges 1995), from which our
mean field %%@
predictions for physical observables
will follow.
Despite their apparent simplicity, these results contain a great
deal of structure, and a fairly careful and non-trivial analysis is required
to extract the universal information contained within them.

First, it is easy to note that there is no sensible solution (with $\Delta >
0$)
of (\ref{mft5}) at $T=0$ for $\widetilde{r} < \widetilde{r}_c$ where
\begin{equation}
\widetilde{r}_c = u \int \frac{d \omega}{2 \pi} \sqrt{|\omega|} \approx \frac{2
%%@
\Lambda_{\omega}^{3/2}}{
3 \pi}
\label{mft6}
\end{equation}
where $\Lambda_{\omega}$ is an upper cut-off in frequency. Clearly the system
is in the
MSG phase for $T=0$, $\widetilde{r} < \widetilde{r}_c$, and a separate ansatz
for $Q$ is
necessary  there, as discussed
elsewhere (Sachdev {\it et al.} 1995). Let us now define
\begin{equation}
r \equiv \widetilde{r} - \widetilde{r}_c,
\label{mft7}
\end{equation}
so that the quantum critical point is at $T=0$, $r=0$. In the vicinity of
this point, our action constitutes a continuum quantum field theory (CQFT)
describing the physics of the system at all energy scales significantly
smaller than $\Lambda_{\omega}$. The ``universal'' properties of the system
are the correlators of this CQFT, and they apply therefore for
$r, T \ll \Lambda_{\omega}$, a condition we assume in our analysis
below. It is also natural to assume that the microscopic coupling
$u \sim \Lambda_{\omega}^{-1/2}$. We shall, however, make no assumptions
on the relative magnitudes of $r$ and $T$.

The general solution of (\ref{mft5}) under the conditions
noted above was described by Sachdev \& Read (1996);
we only present the final result for $\Delta$, which is in the form
of a solvable quadratic equation for $\sqrt{\Delta}$:
\begin{equation}
\Delta + u T \sqrt{\Delta} =
r \left( 1 - \frac{u \Lambda_{\omega}^{1/2}}{\pi} \right)
+ u T^{3/2} \Phi \left(\frac{r}{T} \right),
\label{mft9a}
\end{equation}
where the universal crossover function $\Phi(y)$ is given by
\begin{equation}
\Phi (y) =
\frac{1}{\pi^2} \int_0^{\infty}
\sqrt{s} ds \left[
\log\left( \frac{s}{2 \pi} \right) - \psi \left( 1 + \frac{s+y}{2 \pi} \right)
+ \frac{\pi + y}{s} \right].
\label{mft10}
\end{equation}
with $\psi$ the digamma function.
The following limiting results,
which follow from (\ref{mft10}), are useful for our subsequent analysis:
\begin{equation}
\Phi(y) = \left\{
\begin{array}{cc}
\sqrt{1/2\pi} \zeta(3/2) + {\cal O}(y) & y \rightarrow 0 \\
(2 /3 \pi) y^{3/2} + y^{1/2} + (\pi/6) y^{-1/2}  + {\cal O}(y^{-3/2})
& y \rightarrow \infty
\end{array}
\right.
\end{equation}

\begin{figure}[tb]
\epsfxsize=14in
\centerline{\epsffile{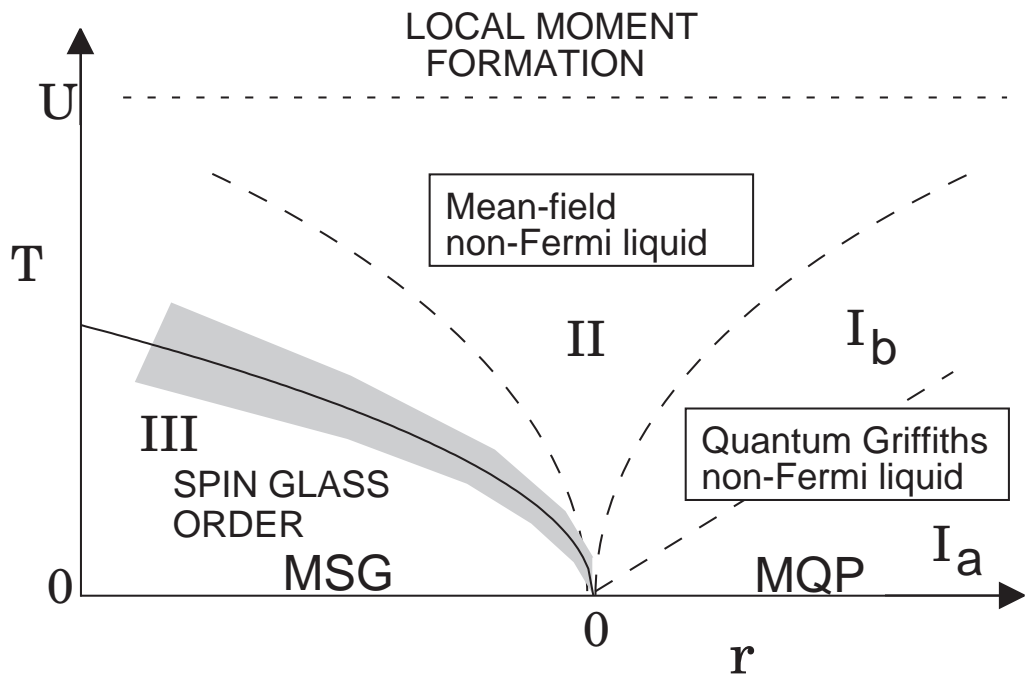}}
\vspace{0.15in}
\caption{Phase diagram of a metallic spin glass as a function of
the ground state tuning parameter $r$ and temperature $T$.
The mean-field theory of local moment formation discussed in
Section~3(b) applies at temperatures $T$ greater than $U$
for all values of $r$.
In the notation
of Fig~\protect\ref{fig1}, the $T=0$ state is a MSG for $r<0$ and
a MQP for $r>0$.
The full line is the only thermodynamic phase transition, and is
at $r = r_c (T)$ or $T = T_c (r)$.
The quantum critical point is at $r=0$, $T=0$,
and is described by a continuum quantum field theory (CQFT).
The dashed lines denote crossovers between different finite $T$ regions
of the CQFT: the low $T$ regions are Ia,Ib (on the paramagnetic side)
and III (on the ordered side), while the high $T$ region (II) displays
``mean-field non-Fermi liquid'' behavior.
The quantum Griffiths precursors of the MSG phase
occur in region I, and now rare regions
lead to certain ``quantum Griffiths non-Fermi liquid'' characteristics
in the thermodynamics.
The crossovers on either side of II, and the spin glass phase boundary
$T_c (r)$, all scale as $T \sim |r|^{z \nu/( 1 + \theta_u \nu)}$;
the boundary between Ia and Ib obeys $T \sim r^{z \nu}$. The mean field
values of these exponents are $z =4$, $\nu = 1/4$, and $\theta_u = 2$.
The shaded region has classical critical fluctuations described
by theories of the type discussed by Fischer \& Hertz (1991)}
\label{fig5}
\end{figure}

The expression (\ref{cross2}),
combined with the results (\ref{mft9a}) and (\ref{mft10})
completely specify the $r$ and $T$ dependence of the dynamic susceptibility
in the MQP phase, and allow us to obtain the phase diagram shown in
Fig~\ref{fig5}.
The crossovers shown are properties of the CQFT
characterizing the quantum critical point. We present below explicit results
for the crossover functions of a number of observables
within the mean field theory.

Before describing the crossovers, we note that
the full line in Fig.~\ref{fig5} denotes the
boundary of the paramagnetic phase at $r = r_c (T)$ (or $T = T_c (r)$).
This is the only line of thermodynamic phase transitions, and its location is
determined by the condition $\Delta = 0$, which gives us
\begin{equation}
r_c (T) =  - u \Phi(0) T^{3/2}~~~\mbox{or}~~~T_c (r) = (-r / u \Phi (0))^{2/3}
\label{mft10a}
\end{equation}

The different regimes in Fig~\ref{fig5} can be divided into two classes
determined
by whether $T$ is ``low'' or ``high''. There are two low $T$ regimes, one for
$r>0$, and the other for $r<0$; these regions display properties of the
non-critical ground states, which were reviewed in Section~\ref{general}.
More novel is the high $T$
region, where the most important energy scale is set by $T$, and
``non-Fermi liquid'' effects associated with the critical ground state
occur. We now describes the regimes in more detail, in turn.

\begin{figure}[tb]
\setlength{\unitlength}{0.8pt}
\begin{picture}(500,230)
\put(20,210){\large\bf LOW T REGION OF CQFT}
\put(50,160){\vector(1,0){400}}
\put(50,160){\line(0,1){20}}
\put(100,170){\large Fermi liquid}
\put(230,160){\line(0,1){20}}
\put(300,170){\large Critical}
\put(440,130){\large $\omega$}
\put(227,140){\large $r^{z \nu}$}
\put(50,140){\large 0}
\put(20,80){\large\bf HIGH T REGION OF CQFT}
\put(50,30){\vector(1,0){400}}
\put(50,30){\line(0,1){20}}
\put(60,40){\large Quantum relaxational}
\put(230,30){\line(0,1){20}}
\put(300,40){\large Critical}
\put(440,0){\large $\omega$}
\put(227,10){\large $u^{z \nu} T^{1+ \theta_u \nu}$}
\put(50,10){\large 0}
\end{picture}
\vspace{0.15in}
\caption{Crossovers as a function of frequency, $\omega$, in the regions
of Fig~\protect\ref{fig5}. The low $T$ region is on the paramagnetic
side ($r > 0$). The quantum Griffiths effects occur in the ``Fermi liquid''
region, making it the ``quantum Griffiths non-Fermi liquid'' of
Fig~\protect\ref{fig5}.}
\label{fig6}
\end{figure}
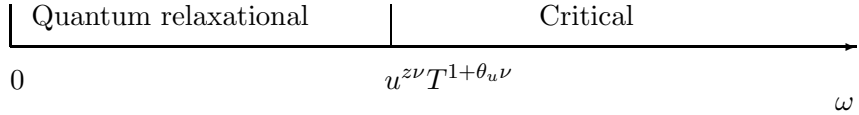

(I) \underline{Low $T$ region above MQP ground state, $T < (r/u)^{2/3}$}
\newline
This is the mean-field ``Fermi liquid'' region, where the leading contribution
to $\Delta$ is its $T=0$ value $\Delta(T) \sim \Delta (0) = r$.
The leading temperature dependent correction to $\Delta$ is
however different in two subregions.
In the lowest $T$ region Ia, $T < r$, we have the Fermi liquid $T^2$
power law
\begin{equation}
\Delta (T) - \Delta (0) = \frac{u \pi T^2}{6 \sqrt{r}}~~~~~~~~\mbox{region Ia}.
\end{equation}
At higher temperatures, in region Ib, $r < T < (r/u)^{2/3}$,
we have an anomalous temperature dependence
\begin{equation}
\Delta (T) - \Delta (0) = u \Phi (0) T^{3/2}~~~~~~~~\mbox{region Ib and II}.
\label{mft11}
\end{equation}
It is also interesting to consider the properties of  region I as a function
of observation frequency, $\omega$, as sketched in Fig.~\ref{fig6}.
At large frequencies, $\omega \gg r$, the local dynamic susceptibility
behaves like $\chi_L^{\prime\prime} \sim \mbox{sgn}(\omega) \sqrt{|\omega|}$,
which is the spectrum of critical fluctuations; at the $T=0$, $r=0$ critical
point,
this spectrum is present at all frequencies. At low frequencies, $\omega \ll
r$,
there is a crossover (Fig~\ref{fig6}) to the characteristic Fermi liquid
spectrum
of local spin fluctuations $\chi_L^{\prime\prime} \sim \omega/\sqrt{r}$.

The present mean-field theory of course does not contain Griffiths effects
coming from rare local fluctuations. These were discussed in Section~\ref{lm},
and as noted there, and are crucial in understanding the low $T$ properties
of the MQP phase. The present mean-field theory captures the behavior of
a ``typical'' region of the sample, but it is the rare regions that dominate
the
thermodynamics and lead to certain ``quantum Griffiths non-Fermi liquid''
characteristics.

(II) \underline{High $T$ region, $T > (|r|/u)^{2/3}$}
\newline
This is the ``mean-field
non-Fermi liquid'' region; unlike region (I), the non-Fermi liquid
effects are now properties of the mean-field theory, and not consequences
of rare fluctuations.
Here temperature dependent
contributions to $\Delta$ dominate over those due to the deviation
of the coupling $r$ from its critical point, $r=0$. Therefore thermal effects
are dominant, and the system behaves as if its microscopic couplings are
at those of the critical ground state. The $T$ dependence in
(\ref{mft11}) continues
to hold, as we have already noted, with the leading contribution now being
$\Delta \approx u \Phi (0) T^{3/2}$.
As in (I), it is useful to consider properties of this region as a
function of $\omega$ (Fig~\ref{fig6}).
For large $\omega$ ($ \omega \gg u T^{3/2}$) we again have the critical
behavior
$\chi_L^{\prime\prime} \sim \mbox{sgn} (\omega ) \sqrt{|\omega|}$; this
critical behavior
is present at large enough $\omega$ in all the regions of the phase diagram.
At small $\omega$ ($ \omega \ll u T^{3/2}$), thermal fluctuations quench the
critical fluctuations, and we have relaxational behavior with
$\chi_L^{\prime\prime} \sim \omega / u^{1/2} T^{3/4}$.
\newline
(III) \underline{Low $T$ region above MSG ground state, $T < (-r/u)^{2/3}$}
\newline
Effects due to the formation of a static moment are now paramount.
As one approaches the spin glass boundary (\ref{mft10a}) from above,
the system enters a region of purely classical thermal fluctuations,
$|T - T_c (r) | \ll u^{2/3} T_c^{4/3} (r)$
(shown shaded in Fig~\ref{fig5}) where
\begin{equation}
\Delta = \left(\frac{r-r_c (T)}{T u} \right)^2
\end{equation}
Notice that $\Delta$ depends on the square of the distance from the
finite $T$ classical phase transition line, in contrast to its linear
dependence, along $T=0$, on the deviation from the quantum critical point
at $r=0$.

We have now completed a presentation of the mean field predictions for the
finite $T$
crossovers near the quantum critical point (Fig~\ref{fig5}), and for the
explicit crossover
functions for the frequency-dependent local dynamic susceptibility
(Fig~\ref{fig6}
and Eqns (\ref{cross2}), (\ref{mft9a}), (\ref{mft10})).
We also summarize here results for a number of
other experimental observables, whose $T$ and $r$ dependences follow from those
of $\Delta$ described above.
\newline
{\it Nuclear relaxation rate $1/T_1$:} (Sengupta \& Georges 1995)
\begin{equation}
\frac{1}{T_1 T} = A^2 \lim_{\omega \rightarrow 0} \frac{\chi_L^{\prime\prime}
(\omega)}{
\omega} = \frac{A^2}{2 \kappa \sqrt{\Delta}},
\end{equation}
where $A$ is determined by the hyperfine coupling.
\newline
{\it Uniform linear susceptibility, $\chi_u$:}
\begin{equation}
\chi_u  = \chi_b - \frac{g}{t \kappa} \sqrt{\Delta},
\end{equation}
where $\chi_b$ is the $T$, $r$- independent background contribution of the
fermions
that have been integrated out, $g$ is a coupling constant.
\newline
{\it Nonlinear susceptibility $\chi_{\rm nl}$:}
\begin{equation}
\chi_{nl} = \frac{u g^2}{4t} \frac{1}{\Delta},
\end{equation}
Notice that $\chi_{nl}$
is proportional to the quantum mechanical interaction $u$, and would vanish in
a
theory with terms associated only with disorder fluctuations.
\newline
{\it Free energy and specific heat:}
\newline
The result for the free energy was given in (\ref{mft5a}), and it needs to be
evaluated along the lines of the analysis carried out above for the crossover
function determining $\Delta$. Such a calculation gives
\begin{eqnarray}
\frac{{\cal F} (T, r) - {\cal F}(T=0,r=0)}{n} =
-\frac{1}{\kappa^2 t} \left[ \frac{2 r \Lambda_{\omega}^{3/2}}{3 \pi} \right.
 &+&
T^{5/2} \Phi_{{\cal F}} \left( \frac{\Delta}{T} \right)
+ \frac{\Lambda_{\omega} \Delta^2}{2 \pi}  \nonumber \\
&+& \left. \frac{(\Delta - r)^2}{2 u} - \frac{4 \Delta^{5/2}}{15 \pi} \right],
\end{eqnarray}
where
\begin{equation}
\Phi_{{\cal F}} (y) = - \frac{2 \sqrt{2}}{3 \pi}
\int_0^{\infty} \frac{ d \Omega}{e^{\Omega}
- 1} \frac{ \Omega (2 y + \sqrt{\Omega^2 + y^2})}{
\sqrt{ y + \sqrt{ \Omega^2 + y^2}}}
\end{equation}
This result for ${\cal F}$ includes non-singular contributions,
smooth in $r$, which form a background to the singular critical
contributions. In region II, the most singular term is the one
proportional to $\Phi_{\cal F}$, and yields a specific heat, $C_v$ (Sengupta \&
Georges
1995):
\begin{equation}
\frac{C_v}{T} = \gamma_b - \frac{\zeta (5/2) }{\sqrt{2 \pi} \kappa^2 t}
\sqrt{T}
\end{equation}
where $\gamma_b$ is a background contribution.
\newline
{\it Charge transport:}
\newline
The consequences of the order parameter fluctuations on charge transport
were explored by Sengupta and Georges (1995). The quasiparticles are
assumed to carry both charge and spin, and they scatter off the
spin fluctuations via an exchange coupling. In the Born approximation,
this leads to a contribution to the quasi particle relaxation rate,
$1/\tau_{qp}$
\begin{equation}
\frac{1}{\tau_{qp}} \propto \int_0^{\infty} \frac{d \Omega}{\sinh(\Omega/T)}
\frac{\Omega}{\sqrt{\Delta + \sqrt{\Omega^2 + \Delta^2}}},
\end{equation}
whose $T$ and $r$ dependence follows from that of $\Delta$.
In region II, $1/\tau_{qp} \sim T^{3/2}$.

\subsection{Application to experiments}
There has been much recent experimental activity on
the transport, thermodynamic and neutron scattering properties
of a number of rare-earth intermetallic compounds which
diplay low $T$ ``non-Fermi liquid'' behavior.
A majority of these systems are not too far from a magnetically
ordered phase of some type: spin-glass ordering has been observed
in Y${}_{1-x}$U${}_{x}$Pd${}_{3}$ (Gajewski {\it et al.} 1996),
La${}_{1-x}$Ce${}_{x}$Cu${}_{2.2}$Si${}_{2}$
(Steglich {\it et al.} 1996), and URh${}_{2}$Ge${}_{2}$
(Sullow {\it et al.} 1997).
However, it is fair to say that a direct relationship between
the properties of these materials and the complicated theoretically
predicted set of crossovers in Fig~\ref{fig5} has not yet been
clearly established. The material La${}_{1-x}$Ce${}_{x}$Cu${}_{2.2}$Si${}_{2}$
has been studied in region II of Fig~\ref{fig5} (Steglich {\it et al.} 1996)
and it appears that initial results for the
resistivity and the specific heat agree well with the mean-field
theoretical predictions
reviewed in Section 4(c).

\section{Conclusions}
In a recent conference on ``non Fermi liquid behavior in metals''
(Coleman {\it et al.} 1996),
two routes to non-Fermi liquid behavior
were discussed: those due to so-called Kondo disorder models,
and those due to proximity to a magnetic quantum phase transition.
These were viewed as competing mechanisms which possibly applied
to different rare-earth intermetallic compounds.
One of the purposes of this article has been to argue
that these mechanisms are really better viewed as different limiting
regimes of the same underlying physics, and that no material
is strictly in one or the other regime. Our basic point is made clear
by a glance at Fig~\ref{fig5}. The Kondo disorder models apply
at low $T$ above the MQP ground state: here rare
local moments with anomalously
low Kondo temperatures appear to dominate and lead to non-Fermi liquid effects.
We have argued here that these effects are best viewed as quantum Griffiths
singularities associated with the transition to the MSG phase (or in
more regular systems, as a transition to some ordered metallic
antiferromagnet), and they are denoted in Fig~\ref{fig5}
as ``quantum Griffiths non-Fermi liquid''.
The other region of non-Fermi liquid behavior appears in region II,
describable as the high $T$ region of the CQFT associated with the
MQP-MSG quantum critical point. Now the non-Fermi liquid behavior
is associated with the behavior of the typical local moment,
and is accessible in mean-field theory; this is denoted
in Fig~\ref{fig5} as ``mean-field non-Fermi liquid''.
Clearly the fluctuation terms which disrupted the mean-field
predictions in region I are also going to be significant here, but there
is no clear understanding of their structure.
A unified theoretical description which includes such corrections
and the crossover to the quantum Griffiths singularities in region I
is lacking, and should be an important focus of future theoretical work.

Finally, a few speculative remarks on the metal-insulator transition,
which has been the focus of so much theoretical attention in the last
decade. One of the stumbing block in the analysis of the transition
from the metal (MQP) to the insulator (IQP) has been the apparent
run-away flow of the ``triplet interaction amplitude'' to infinity
(Belitz \& Kirkpatrick 1994). We believe this runaway flow is the
signal within the weak-disorder perturbation theory of the local
moment instability reviewed here. We have also argued that this instability
is really a precursor of the MQP-MSG transition, and is therefore
incidental to the MQP-IQP transition (it should also be noted
that Belitz \& Kirkpatrick (1996) have argued that the runaway
flow is related to a ferromagnetic phase). The correct theory of the
latter transition should ``factor out'' these local moments in some way.
How this may be consistently done, is not understood. Clearly questions related
to the interplay and separation of critical and quantum Griffiths singularities
discussed above, arise here too.

\section*{Acknowledgements}
I thank T.~Senthil for a recent collaboration (Senthil \& Sachdev 1997) and for
helpful remarks which
greatly influenced the perspective of this article.
The research was supported by the U.S. National Science Foundation
under grant number DMR-96-23181.

\end{document}